\documentclass[a4paper,11pt,preprint,manuscript]{cpbtex}
\usepackage{graphicx}
\usepackage{float}
\usepackage{units}
\allowdisplaybreaks

\begin{document}

\title{Epitaxially strained SnTiO$_{3}$ at finite temperatures
\thanks{Project supported by the National Natural Science Foundation of China
	(Grant No.~11574246, 51390472, U1537210, and 11564010),
the National Basic Research Program of China (Grant No. 2015CB654903),
the Natural Science Foundation of Guangxi (GA139008),
and the "111" Project (Grant No. B14040).
 }}

  \author{
  Dawei Wang$^{\rm a)}$\thanks{Corresponding author. E-mail:dawei.wang@mail.xjtu.edu.edu}, \and
  Laijun Liu$^{\rm b)}$\thanks{Corresponding author. E-mail:2009011@glut.edu.edu}, \and
  Jia Liu$^{\rm c)}$,\and Nan Zhang$^{\rm d)}$, \and Xiaoyong Wei$^{\rm d))}$\\
  \and
  $^{\rm a)}$School of Microelectronics and State Key Laboratory for Mechanical Behaviour of Materials,\\
  Xi'an Jiaotong University, Xi'an 710049, China \\
  \and
  $^{\rm b)}$College of Materials Science and Engineering,\\
  Guilin Univeristy of Technology, Guilin 541004, China
  \and
  $^{\rm c)}$State Key Laboratory for Mechanical Behavior of Materials \&\\
  School of Materials Science and Engineering, Xi'an Jiaotong University, Xi'an 710049, China
  \and
  $^{\rm d)}$Electronic Materials Research Laboratory-Key Laboratory of the Ministry of Education and\\
  International Centre for Dielectric Research, Xi'an Jiaotong University, Xi'an 710049, China
}
  \date{\today}
\maketitle

\begin{abstract}
Combining the effective Hamiltonian approach and direct \emph{ab initio}
computation, we obtain the phase diagram of SnTiO$_{3}$ with respect to
epitaxial strain and temperature, demonstrating the complex features of the
phase diagram and providing insight into this system, a presumably simple
perovskite. Two triple points, as shown in the phase diagram, may be exploited
to achieve high-performance piezoelectric effects. Despite the inclusion of the
degree of freedoms related to oxygen octahedron tilting, the ferroelectric
displacements dominate the structural phases over the whole misfit strain range.
Finally, we show SnTiO$_{3}$ can change from hard to soft ferroelectrics with
the epitaxial strain.
\end{abstract}

\textbf{Keywords:} SnTiO$_3$, phase diagram, epitaxial strain

\textbf{PACS:} 77.80.-e, 77.84.-s, 81.30.Bx

\section{Introduction}

Piezoelectricity is a phenomenon in certain materials that strain
and electric polarization can induce and/or influence each other.
Ferroelectric materials, which are inherently piezoelectric, can produce
an electric polarization proportional to the load, in response to
an applied mechanical strain. Similarly, such materials will produce
a mechanical deformation (strain) in response to an applied voltage.
Switchable polarization makes ferroelectrics a critical component
in memories, actuators, electro-optic devices, and potential candidates
for nanoelectronics \ucite{Nature2015}.

In recent years, it has been found that materials of high-performance
piezoelectricity are often associated with morphortropic phase boundary (MPB),
with examples including Pb(Zr,Ti)O$_{3}$\,\ucite{Jaffe} and
(K,Na)NbO$_{3}$-LiTaO$_{3}$-LiSbO$_{3}$\,\ucite{Saito}, or triple points, e.g.
in (Ba,Ca)(Zr,Ti)O$_{3}$\ucite{PRL2009}. Engineering solid solutions to a
certain composition can create phase boundaries and tricritical points, where
the crystal structure changes abruptly, inducing maximal piezoelectric
properties. Three important situations have been extensively studied: (i) MPB in
pure perovksites that separates regions of the tetragonal from the rhombohedral
symmetry \ucite{JAP1954}; (ii) MPB formed in perovskites dissoluted with a small
amount of non-perovskite-structured materials that can cause lattice distortions
\ucite{Saito} and grain boundary effects \ucite{Zeng2018}; (iii) Regions close
to a triple point where cubic paraelectric phase (C), ferroelectric rhombohedral
(R), and tetragonal (T) phases meet \ucite{PRL2009}. \,\, In addition to MPB,
epitaxial thin-film growth, which introduces intrinsic lattice strain, has
matured as another important method to design desired ferroelectric materials,
importan for highly integrated design and intelligent control technology
\ucite{Wessels2007,Ramesh2007}. Strain engineering, widely adopted, can tune the
large $2p$-$3d$ charge hybridisation between the strongly correlated $3d$
electrons in transition metal ions and the $2p$ electrons of oxygens
\ucite{Hwang2012,Yamada2004,Ohtomo2004,Spaldin2005}. For instance, both
compressive and tensile strains increase the Ni $3d$ band width and favour the
metallic phase in NdNiO$_{3}$ \ucite{Wang2015}. In addition, substrate clamping
will force the temperature dependence of in-plane lattice constants of the grown
films to follow that of the substrates, which may lead to unexpected phase
transitions and domain formation \ucite{He2004,Jiang2014}. Therefore, strain
constraint can even introduce muti-phase coexistence in thin films, which makes
it an attractive method to fine tuning properties of films.

Nowadays commonly used high-performance piezoelectric materials, including
PbTiO$_{3}$ and Pb(Mn,Nb)O$_{3}$, contain hazardous lead (Pb). Since Pb is
harmful to environment and human health, lead-free ferroelectric materials are
highly desired. Many lead-free materials are based on
(Bi$_{0.5}$,Na$_{0.5}$)TiO$_{3}$,\textbf{ }(K,Na)NbO$_{3}$ or BaTiO$_{3}$,
however, their performance is still sub-optimal compared to Pb-containing
materials \ucite{Wu2015}. Since Sn and Pb belong to the same family, SnTiO$_{3}$
is expected to achieve high-performance piezoelectricity with environmentally
benign elements \ucite{Armiento2009}. Indeed, SnTiO$_{3}$ has large polarization
and large axial ratio \ucite{Lebedev2009,Parker2011}, even larger than
PbTiO$_{3}$, makeing it a promising candidate. But for various reasons,
SnTiO$_{3}$ bulk material is hard to prepare, since Sn$^{2+}$ can
easily become Sn$^{4+}$, and Sn is prone to enter into the B site (where the Ti
ion stays) due to its small ionic radius. However, researchers continued to look
for opportunities to exploit the remarkable properties of SnTiO$_{3}$. For
instance, researchers have considered Sn-doped BaTiO$_{3}$ \ucite{Xie2009,add1},
Bennett\emph{ et al} considered Sn(Al$_{0.5}$,Nb$_{0.5}$)O$_{3}$\ucite{Bennett},
while Suzuki\emph{ et al} obtained Sn-doped SrTiO$_{3}$ \ucite{Suzuki2012}, and
Laurita investigated (Sr,Sn)TiO$_{3}$ and (Ba,Ca,Sn)TiO$_{3}$
\ucite{Laurita2015}. Recently, Agarwal et al \ucite{Agarwal2018} obtained
perovskite phase SnTiO$_3$ with the atomic-layer deposition technique. This is
an important breakthrough that will excite more work on SnTiO$_{3}$.

In addition to experimental work, there are also many theoretical investigations
on SnTiO$_{3}$
\ucite{Armiento2009,Parker2011,Taib2013,Taib2014,Uratani2008,Zhang2013,Zhang2014}.
However, most of previous theoretical investigations are based on direct
\emph{ab initio} computation, which cannot provide information regarding
SnTiO$_{3}$ at finite temperatures. Therefore, important questions remain
unanswered. For instance, what are the conditions for the existence of different
ferroelectric phases? How is the phase diagram of SnTiO$_{3}$ like? Can such a
seemingly simple perovskite (SnTiO$_{3}$ is not solid solution and no doping is
applied) possess complex features? In this work, we will focus on the finite
temperature properties of epitaxially strained SnTiO$_{3}$ and address these
questions. We note that such information will be useful for the fabrication of
SnTiO$_{3}$ bulk, or the growth of SnTiO$_{3}$ film, and engineering
SnTiO$_{3}$-containing ferroelectric materials. Since SnTiO$_{3}$ bulk is not
available, there is not much \emph{a priori} information at finite temperatures
that can be used in this work. While our computational results are less
convincing without the support from experiments, it demonstrates the value of
theoretical and numerical work, i.e., their power to predict something unknown
-- this is one of the reasons that motivated this investigation.

\section{Method \label{sec:Method}}

To fully understand SnTiO$_{3}$, knowing all its possible phases under various
conditions is desired. However, it is not trivial to achieve this goal. For
instance, the direct \emph{ab initio} approach usually provides us with the
structural phase of local energy minimum (in contrast to the most stable phase
of global energy minimum), which is exacerbated by the 0 K assumption adopted.
The application of this approach often requires the comparison of many different
phases that may not be able to cover all phases. To address this problem, here
we adopt the first-principles based effective Hamiltonian approach and
Monte-Carlo (MC) simulations, which were developed exactly to address such
challenges. To use this approach, it is necessary to compute tens
of coefficients appearing in the effective Hamiltonian using direct \emph{ab
  initio} computation before carrying out the MC simulations.

The
effective Hamiltonian used here was originally developed in Ref. \cite{Zhong1994},
which incorporates the coupled dynamics of the soft mode, strain,
and dipole. Its internal energy is given by 
\begin{align*}
E= & E^{\textrm{FE}}\left(\left\{ \mathbf{u}_{i}\right\} ,\left\{ \mathbf{v}_{i}\right\} ,\eta_{H}\right),
\end{align*}
where $\mathbf{u}_{i}$ is the local soft-mode in unit cell $i$,
and is proportional to the local electric dipole moment in that cell
when multiplied by Born effective charge. We note that $\mathbf{u}_{i}$
here is located on the A-site (where Sn stays), and represents the
collective motion of Sn, Ti, and O atoms inside one unit cell. The
$\mathbf{v}_{i}$ are Sn-centered local displacements related to the
inhomogeneous strain inside each unit cell. $\eta_{H}$ is the homogeneous
strain tensor. The energy terms and associated parameters can be found
in Ref. \cite{Ye2016}. In this work, we extended the effective Hamiltonian
and expanded the local energy term to 8th order (similar to Ref.\cite{Nishimatsu})
in order to describe the internal energy more precisely \ucite{Vanderbilt2001}.
In addition, the antiferro-distortive (AFD) oxygen octahedron tilting
is also considered with the energy term becoming\ucite{Kornev2006}
\begin{align*}
E= & E^{\textrm{AFD}}\left(\left\{ \boldsymbol{\omega}_{i}\right\} ,\left\{ \mathbf{u}_{i}\right\} ,\left\{ \mathbf{v}_{i}\right\} ,\eta_{H}\right),
\end{align*}
where the new variable $\boldsymbol{\omega}_{i}$ represents the oxygen tilting
on the unit cell $i$, which centres on the B-site atom (i.e., Ti). The
energy term $E^{\textrm{AFD}}$ also includes the couplings between
AFD, and all the other dynamical variables. The associated parameters
in $E^{\textrm{AFD}}$ are obtained in a similarly way as in Ref.
\cite{Ye2013,Ye2016}. The effective Hamiltonian approach is a well
established methodology that has been developed since 1994 \ucite{King-Smith1994,Zhong1994}
and similar to direct \emph{ab initio} methods, it has been used in many investigations, e.g.,
to predict new structural phases of perovksites \ucite{PMN-2015,Jiang2015}.
The most appealing features of this approach are (i) It can
often lead us to the most stable structural phase of global energy minimum; (ii)
It can produce finite temperature properties, which explains why
we adopt this approach in this work.

With the effective Hamiltonian, we perform MC simulations on a
$12\times12\times12$ supercell, containing 8640 atoms. The system is set to meet
the periodic boundary condition along the (pseudo) $x,y,z$ directions and subject
to various epitaxial strains. In each simulation at a given epitaxial misfit
strain $s$, we gradually cool down the system from 2500 K to 5 K. For each
temperature, we carry out 160,000 MC steps to obtain averaged physical
quantities -- most importantly -- the supercell average of local mode, which can
be used to determine the symmetry of the system and the approximate positions of
each atom in it.

We have also performed direct \emph{ab initio} computation to corroborate
MC simulation results. For this purpose,
the open source ABINIT software package \ucite{Gonze2002} is
used along with the local density approximation (LDA) \ucite{Perdew1992}
and the projector-augmented-wave (PAW) method \ucite{Blochl1994}.
We use the pseudo-potentials implemented in the GBRV package \ucite{Garrity2014},
and the Sn\emph{ 4d 5s 5p}, Ti \textit{3s 3p 4s 3d}, and O \textit{2s
2p} orbitals are treated as valence orbitals. For convergence, we
have chosen the cut-off energy (\texttt{ecut}) to be 25 Hartree (1
Hartree = 27.211 eV) for plane wave expansion, and fine grid the
cut-off energy (\texttt{pawcutdg}) is selected to be 
50 Hartree. In addition, $k$-point sampling of
$6\times6\times6$ Monkhorst-Pack grid \ucite{Monkhorst1976} was
used. The atomic coordinates are relaxed until all atomic-force components
are smaller than $10^{-5}$ Ha/Bohr, and the cell size and shape are
varied until all stress components are below $10^{-7}$ Hatree/Bohr$^{3}$.

Both the MC simulations and the direct \emph{ab initio} computations deal with
the strained bulks (in contrast to ultrathin two-dimensional films) given the
periodic boundary condition used. However, many properties of the strained films (as
long as they are \emph{not} just a few atomic layers thick) can be inferred from
such calculations. Comparing to our previous work \ucite{Ye2016}, the present investigation
develops in three directions: (i) We now include the new degree of
freedom, i.e., oxygen octahedron tilting (often called antiferrodistortive
rotations, AFD), which is ignored in the previous work \ucite{Ye2016}.
(ii) To this end, we computed many new essential parameters for the
effective Hamiltonian. While the current work build on the previous
one, it has substaintially extended the effort to fully simulate SnTiO$_{3}$.
(iii) Here we considere the effects of epitaxial strain and show
that SnTiO$_{3}$ has a rather complex phase diagram, containing interesting
phase boundaries and triple points.

\section{Results\label{sec:Results}}

In order to obtain the phase diagram, we have performed MC simulations
for misfit strain between $s=0$ and $s=2\%$ with a step size of
$\Delta s=0.125\%$. For each misfit strain $s$, the system is gradually
cooled down from $\sim2500$\,K to $5$\,K and its evolution
with temperature is observed. The simulation results are then summarized
to form the phase diagram.

\subsection{Phase transition and phase diagram}

\begin{figure*}
\begin{centering}
\noindent\includegraphics[width=12cm]{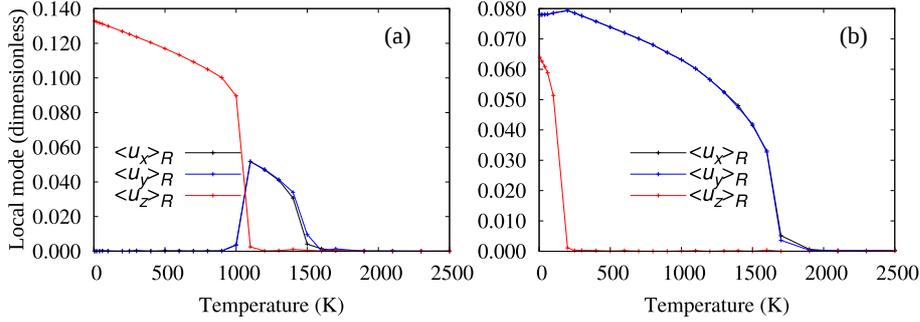}
\end{centering}
\caption{Local mode versus temperature at two given strain misfit, $s=$0.375\%
(Left panel) and $s=$1\% (Right panel). \label{fig:local_mode} }
\end{figure*}

We first obtain the supercell averaged local mode versus temperature
of SnTiO$_{3}$ under various tensile epitaxial strains, which is
expected to stabilize SnTiO$_{3}$\ucite{Parker2011}.
Such information enables us to find the evolution of structural phases
with respect to temperature and epitaxial strain, and more importantly,
the phase transition temperatures of the system. Figure \ref{fig:local_mode}
shows the results with misfit strain $s=$0.375\% and $s=$1\%.
Note that we have relaxed the cubic phase ($Pm\bar{3}m$) SnTiO$_{3}$
using ABINIT (with settings specified in Sec. \ref{sec:Method}\quad) to
obtain the lattice parameter $a=7.312$ Bohr, which is used as the
reference value for specifying the misfit strain.

For the smaller epitaxial strain $s=0.375\%$, the system adopts $P4mm$
phase for temperature $T\lesssim1000$ K, ($P_{x}=P_{y}=0,\thinspace P_{z}>0$),
as the temperature increases it undergoes a phase transition to become
the $Amm2$ phase ($P_{x}=P_{y}>0,\thinspace P_{z}=0$), and eventually
the cubic phase for $T\gtrsim1500$\,K. The whole process
is somewhat similar to that of BaTiO$_{3}$ and (Na$_{0.5}$K$_{0.5}$)NbO$_{3}$,
where several structural phases are involved in phase transitions.
Here, however, a rather drastic change happens at around 1000 K as
the polarization rotates from out-of-plane to the in-plane configuration.
For the larger epitaxial strain ($s=1\%$), there are also two phase
transitions. At low temperature ($T\lesssim197$\,K), it adopts the
$Cm$ ($P_{x}=P_{y}>0,\thinspace P_{z}>0$) phase, which changes to
the $Amm2$ phase as the temperature increases, and finally become
paraelectric at $T\simeq1750$\,K. This phase transition temperature
is high comparing to other typical ferroelectric materials, e.g.,
PbTiO$_{3}$ (763 K) and BiFeO$_{3}$(1100 K)\ucite{Liu2010}. The reason is likely
due to the strong dipole-dipole interaction inside SnTiO3, noting
that the spontaneous polarization of SnTiO$_{3}$ (1.32 C/m$^{2}$)
\ucite{Ye2016} is even larger than the strained BiFeO$_{3}$ (1.30
C/m$^{2}$) \ucite{Zhang2011}.

\begin{figure}[h]
\begin{centering}
\includegraphics[width=8cm]{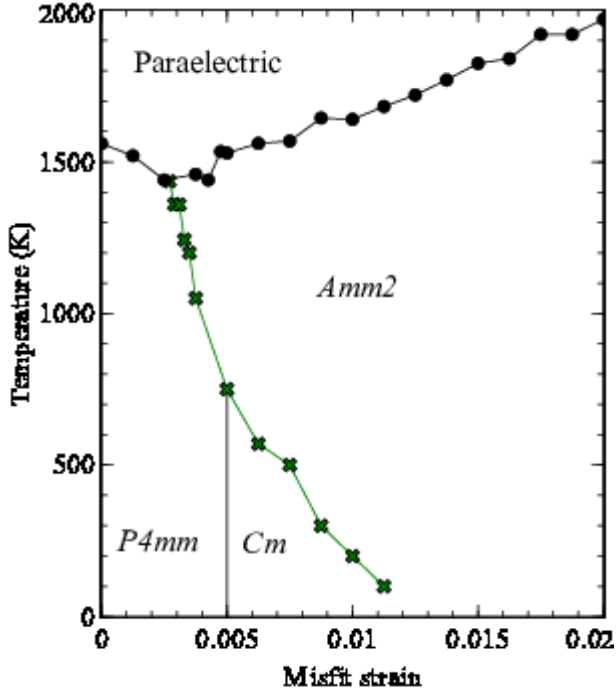}
\end{centering}
\caption{Phase diagram of epitaxially strained SnTiO$_{3}$. \label{fig:phase-diagram}}
\end{figure}

It is worth noting that no correlated AFD was observed for the epitaxial strains
investigated here, consistent with previous known results \ucite{Parker2011}
(also see Sec. \ref{sec:Discussion}\quad). Figure \ref{fig:local_mode} indicates
that four phases (paraelectric $Pm\bar{3}m$, orthorhombic $Amm2$, tetragonal
$P4mm$, and monoclinic $Cm$) can all exist in SnTiO$_{3}$ at proper temperature
and epitaxial strain. Without strain constraint or at tiny strain ($s<0.5\%$,
see Fig. \ref{fig:phase-diagram}), the system adopts the $P4mm$ phase
($P_{z}>0$,$P_{x,y}=0$) at low temperature, and only has one phase transition
($P4mm$ to paraelectric) as temperature increases. Interestingly, Figure
\ref{fig:local_mode}(b) shows a region ($T\leq200$ K) that corresponds to the
$M_{B}$ phase (belong to space group $Cm$) \ucite{Vanderbilt2001,ZhangN2014},
and for smaller $s$ (e.g., at $s=0.75\%$), we have observed that
$P_{z}>P_{x}=P_{y}$, which is also $Cm$, but corresponds to the $M_{A}$ phase
\ucite{ZhangN2014}. Moreover, in Figure 1(a), both the $M_{A}$ and $M_{B}$
phases exist between the T ($P4mm$) and O ($Amm2$) phases. The phase transition
sequence resembles the local structure evolution of Pb(Ti$_{1-x}$,
Zr$_{x}$)O$_{3}$ with increasing $x$ \ucite{Lu2015}. Such monoclinic region has
been shown to play critical role in high-performance piezoelectric materials
\ucite{Liu2017}.

We now turn to the phase diagram of SnTiO$_{3}$ with respect to temperature
and epitaxial strain. First, for all the investigated
misfit strains (up to 2\%), Fig. \ref{fig:phase-diagram} shows that
SnTiO$_{3}$ undergoes a paraelectric to ferroelectric phase transition
at rather high temperature ($T_{C}>1400$\,K). The
high $T_{C}$ is likely due to the large intrinsic spontaneous polarization
discussed previously \ucite{Ye2016}. Second, $T_{C}$
initially decreases with $s$, reaching a minimum, and then increases
again \ucite{Schlom2007}, the lowest point corresponds to a triple
point separating the paraelectric phase and the other two ferroelectric
phases (tetragonal $P4mm$ and orthorhombic $Amm2$), similar to what
happens in ultrathin PbTiO$_{3}$ films \ucite{Jiang2014}. We note
that the existence of the $Cm$ phase is consistent with results obtained
in Ref. \cite{Parker2011}. The abrupt transition
from the $P4mm$ to the $Cm$ phase (with respect to the misfit strain)
happens in a slender region ($<0.125\%$) represented
by a straight line in Fig. \ref{fig:phase-diagram} due to the limit
of $\Delta s$ used in simulations. 

The first triple point in Fig. \ref{fig:phase-diagram}, where one paraelectric
cubic phase and two ferroelectric phases converge, contains potential giant
piezoelectric effects. For instance, such triple point was found and exploited
in binary compounds including
(Ba$_{0.7}$,Ca$_{0.3}$)TiO$_{3}$-Ba(Zr$_{0.2}$,Ti$_{0.8}$)O$_{3}$ (BCZT)
\ucite{PRL2009}, Ba(Sn$_{0.12}$,Ti$_{0.88}$)O3-$x$(Ba$_{0.7}$,Ca$_{0.3}$)O$_{3}$
\ucite{Xue2011}, and other BaTiO$_{3}$-derived systems \ucite{JPD2012}. More
importantly, the three ferroelectric phases ($P4mm$, $Amm2$, and $Cm$) converge
to a second triple point. There are three boundaries around this point. The
boundary at $s=0.5\%$ separates the $P4mm$ and the $Cm$ phase, resembling the
MPB seen in Pb$\left(\textrm{Zr}_{1-x},\textrm{Ti}_{x}\right)$O$_{3}$, in which
two different structural phases exist with a buffer region at $x\simeq0.48$. In
Pb$\left(\textrm{Zr}_{1-x},\textrm{Ti}_{x}\right)$O$_{3}$ the monoclinic phase
serves as a bridge between the higher symmetry tetragonal phase (with
$\left[001\right]$ polarization) and the rhombohedral phases (with
$\left[111\right]$ polarization ) \ucite{Kornev2006}. In this connecting phase,
the polarization can align anywhere on the $\left\{ 110\right\} $
mirror plane between the pesudocubic $\left[111\right]$ and $\left[001\right]$
directions, giving rise to the high piezoelectric response
\ucite{Kornev2006,Guo2000,Cox2001}.

Here in SnTiO$_{3}$, the pronounced existence of the $Cm$ phase and the second
triple point, which bridges the tetragonal phase (with
$\left[001\right]$ polarization) and the orthorhombic phase (with
$\left[110\right]$ polarization), may also enable high performance, following
the pattern of the universal phase diagram discussed in Ref.
\cite{Guo2000,Cox2001}. The reason is similar to that of
Pb$\left(\textrm{Zr}_{1-x},\textrm{Ti}_{x}\right)$O$_{3}$: in this region,
polarization anisotropy nearly vanishes and thus polarization rotations are easy
\ucite{ZhangN2014}. Many alkaline niobate perovskites show a polymorphic phase
transition \ucite{MaterialsLetters2012} between the tetragonal phase and the
orthorhombic phase, similar to what happens in Fig. \ref{fig:phase-diagram}. The
polymorphic behaviour can also lead to high piezoelectricity due to the
instability with respect to polarization rotation \ucite{Wada1999}. However, the
temperature stability of their piezoelectric properties for alkaline niobate
perovskites is not as good as
Pb$\left(\textrm{Zr}_{1-x},\textrm{Ti}_{x}\right)$O$_{3}$. In addition, the
large anisotropy along the whole polymorphic boundary line leads to a larger
energy barrier between the two polarization states (tetragonal and
orthorhombic), preventing possible polarization rotations because the phase
coexistence results from the diffusive tetragonal to orthorhombic phase
transformation. \ucite{PRL2009} In SnTiO$_{3}$, unlike the polymorphic phase
transition, the $Cm$ phase is associated with a triple point, where a low energy
barrier between two ferroelectric phases ($P4mm$ and $Amm2$) may exist that
facilitates the polarization rotation and lattice distortion, leading to high
performance. Unfortunately, this triple point is not at room temperature for
pure SnTiO$_{3}$ (which is at $\sim750$ K as shown in Fig.
\ref{fig:phase-diagram}), and may need to be tuned (e.g by doping). For
instance, following the lessons from BCZT, it may be possible to use Pb to
substitute Sn and/or Zr to substitute Ti. In this way, SnTiO$_{3}$ can be taken
as the matrix material for designing high performance piezoelectric materials.

\begin{figure*}
\begin{centering}
\noindent\includegraphics[width=12cm]{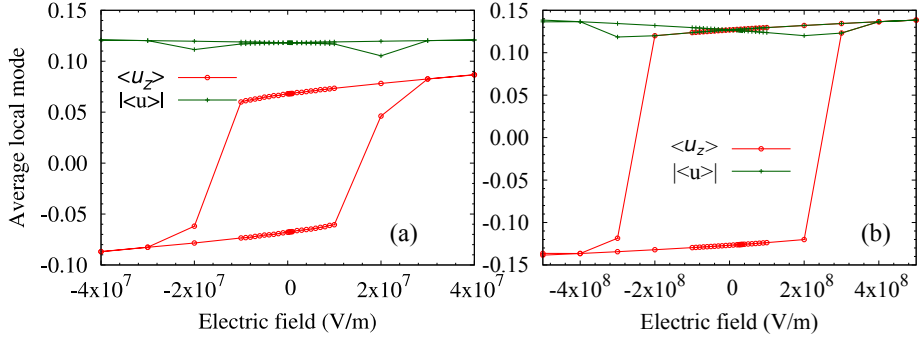}
\end{centering}
\caption{Hysteresis loop of SnTiO3 at $s=0.75\%$ (a) and $s=0\%$ (b) at 300
K. The electric field is applied along the $z$-axis. \label{fig:hyst-loop}}
\end{figure*}

To show that the polarization in the $Cm$ phase can easily rotate, we also
obtained the hysteresis loop at different epitaxial strains. As Fig.
\ref{fig:hyst-loop} shows, the $Cm$ phase has a strong effect on the hysteresis
loop of SnTiO$_{3}$. When the $Cm$ phase exists {[}Fig.
  \ref{fig:hyst-loop}(a){]} at $s=0.75\%$, the coercive field is
$\sim1.6\times10^{7}$ V/m, reduced by a factor of 10 compared to the result
obtained when $s=0\%$ ($>2\times10^{8}$V/m, see Fig. \ref{fig:hyst-loop}(b)). In
analogous to magnetic materials, by applying a proper strain (e.g., $0.75\%)$,
SnTiO$_{3}$ becomes ``soft'' ferroelectrics. Interestingly, in the whole
process, the magnitude of the polarization $\boldsymbol{P}$ is approximately a
constant {[}dark green line in Fig. \ref{fig:hyst-loop}{]}, making the whole
process a rotation as well as switching of polarization (albeit a sudden
rotation), consistent with previous computations \ucite{Zhang2013,Zhang2014}.

\subsection{Direct \emph{ab initio} computation}

\begin{figure}[h]
\begin{centering}
\includegraphics[width=8cm]{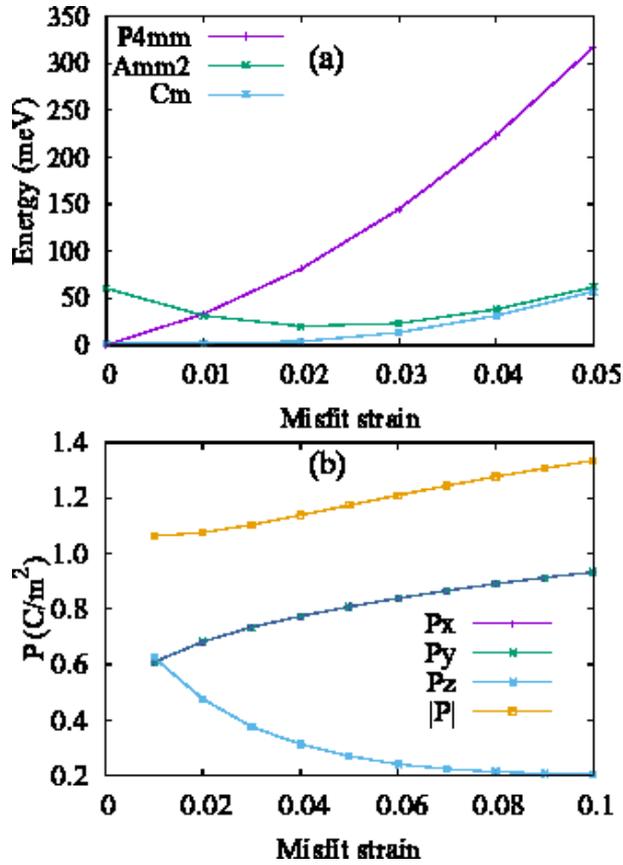}
\end{centering}
\caption{(a) Energy and polarization versus strain for three different phases
$Amm2$, $P4mm$, and $Cm$. At $s=0\%$; (b) The polarization ($P_{x}=P_{y}$
and $P_{z}$) versus strain for the $Cm$ phase. \label{fig:Energy-and-polarization}}
\end{figure}

To corroborate results obtained from effective-Hamiltonian-based computations,
we also obtained numerical results from direct \emph{ab initio} computation.
Figure \ref{fig:Energy-and-polarization}(a) shows the energy versus
strain for the $P4mm$, $Amm2$, and $Cm$ phases. These phases all
appear in the phase diagram of SnTiO$_{3}$ (Fig. \ref{fig:phase-diagram}).
For a large range of epitaxial strain, the $Cm$ phase has the lowest
energy. At $s=1\%$, the energy of the $Cm$ phase is approximately
$29.4$ meV lower than that of $P4mm$ or $Amm2$. Such results indicate
that the $Cm$ phase can be the ground state for a restricted range
of strain, which supports our effective Hamiltonian results. 
The $Cm$ phase appears around $s=0.5\%$ is likely because 
at this point the three phases ($Cm$, $P4mm$, and $Amm2$)
have similar energies, and macroscopically the average of $P4mm$
and $Amm2$ also give rise to the $Cm$ phase. On the other hand,
when $s>1\%$, $P4mm$ is no longer an option as its energy becomes
much higher than the other two. As the energy difference between $Amm2$ and $Cm$ continue
to be smaller with the misfit strain, the $Cm$ phase can only exist at low temperatures,
and eventually disappears. The $Cm$ phase rarely
appear in pure perovskites, whether it is epitaxially strained or
not, and SnTiO$_{3}$ seems to be an important exception. Figure \ref{fig:Energy-and-polarization}(b)
plots the polarization versus the misfit strain for the $Cm$ phase, which is
similar to the results obtained in Ref. \cite{Parker2011}, showing
that $P_{z}$ increases while $P_{x,y}$ decreases with increasing
epitaxial strain. We finally note that, despite many attempts, no
structural phases involving AFD tilting were found to
be the ground state. For $0\%<s<5\%$, structural phases without AFD
are consistently more stable in terms of energy.

\section{Discussion\label{sec:Discussion}}

In Sec. \ref{sec:Results} \quad we have shown that SnTiO$_{3}$ has $P4mm$,
$Amm2$ and $Cm$ phases under different conditions. However, it is
hard to prove that SnTiO$_{3}$ can only has these three phases
for the misfit strain we have investigated. This point is further
discussed below in Sec. \ref{subsec:Seeking-nNew-phases}.\quad Moreover,
we will compare the MC results to direct \emph{ab initio} results
and focus on the $Cm$ phase in Sec. \ref{subsec:The-Cm-phase}.

\subsection{Seeking new phases \label{subsec:Seeking-nNew-phases}}

Phonon calculations have shown that for SnTiO$_{3}$ the AFD-related
modes are also unstable \ucite{Parker2011}. Considering this fact,
it is rather surprising that the MC simulation results are not
able to identify new phases involving oxygen octahedron tilting, which
is a pity and an important lesson. 

We had intentionally played with the coefficients in the
effective Hamiltonian and performed additional MC simulations in order
to suggest new phases involving AFD. The simulations results indeed
generated a few AFD-related candidates (e.g., the $Ima2$, $Imma$,
and $I4cm$ phases). However, direct \emph{ab initio} calculations do not
validate them as energy ground states, consistent with the fact that Parker
\emph{et al} did not propose any AFD related phases although their calculations
have shown strong AFD instability \ucite{Parker2011}. Therefore, our results strongly
suggest that SnTiO$_{3}$ may not have AFD-related phases as ground
state for the misfit strain of $s<5\%$. 

The fact that AFD-related phases do not appear is most likely due to
the strong competition from the polar local mode, which is responsible
for the polarization in SnTiO$_{3}$. For ferroelectric materials,
AFD is usually adverse to the development of polarization.
One term in the effective Hamiltonian specifically represents such
an effect, which is $D\boldsymbol{\omega}^{2}\boldsymbol{u}^{2}$
\ucite{Wang2011}, where the sign of the coupling coefficient ($D$)
will determine how strong the competition (or in rare cases the cooperation)
between the local mode $\boldsymbol{u}$ (related to polarization)
and $\boldsymbol{\omega}$ (related to AFD). In SnTiO$_{3}$, apparently
the ferroelectric local mode (which are responsible for the $P4mm$, $Amm2$,
and $Cm$ phases) dominate the system.

\subsection{The \emph{Cm} phase \label{subsec:The-Cm-phase}}

The results from \emph{ab initio} computations in Fig. \ref{fig:Energy-and-polarization}\quad
show that the \emph{Cm} phase has the lowest energy beyond $s=0$ among
the $P4mm$, $Cm$ and $Amm2$ phases. On the other hand, the phase
diagram obtained from MC simulations shows that SnTiO$_{3}$
exhibits the $Cm$ phase in a much smaller misfit strain range. The
most likely reason of this discrepancy is that the lone pair on Sn$^{2+}$ can
cause the extra out-of-plane polarization that is not well accounted
for in the effective Hamiltonian, where the polarization is closely
related to the $\Gamma$-point (of the cubic phase) unstable polar
mode \ucite{Zhong1994}. 

For SnTiO$_{3}$ that mode is
$u_{x,y,z}=(\xi_{\textrm{Sn}},\xi_{\textrm{Ti}},\xi_{\textrm{O}_{allel}},\xi_{\textrm{O}_{\perp}},\xi_{\textrm{O}_{\perp}})
=\left(0.534,0.169,-0.411,-0.508,-0.508\right)$ (normalized), which
approximately represents the ion displacements in SnTiO$_{3}$ and specifies how
spontaneous polarization develops. However, some anomaly happens for SnTiO$_{3}$
as can be seen from the direct \emph{ab initio} computation. For instance, at
$s=1\%$, along the $z$ direction the ion displacements are
$u_{z}^{\prime}=\left(0.847,0.352,-0.049,-0.27,-0.27\right)$ while along the
$x,y$ directions,
$u_{x,y}^{\prime}=\left(0.781,0.259,-0.150,-0.387,-0.387\right)$. It is
important to note that $u^{\prime}$ has a much larger weight on Sn$^{2+}$, which
is not reflected in $u$. In fact, this is a known issue during the development
of the effective Hamiltonian approach as discussed in Ref. \cite{King-Smith1994}
where it was pointed out that for KNbO$_{3}$ and PbTiO$_{3}$ there is large
difference between the experimental and theoretical local modes. Here,
SnTiO$_{3}$ has the same problem while the difference between PbTiO$_{3}$
and SnTiO$_{3}$ is discussed in detail in Ref. \cite{Pitike2015}. In principle,
for SnTiO$_{3}$ it is possible to tune the values of $u$ (or adding an extra
polar mode) in the effective Hamiltonian to alleviate or fix this issue. But
such a move will involve many more cumbersome calculations of coupling
coefficients, making the approach more complicated. This issue again shows the
complexity of SnTiO$_{3}$ despite its simple composition.

\section{Conclusion}

Using effective Hamiltonian based Monte Carlo simulations, we have
investigated epitaxially strained SnTiO$_{3}$, found their structural
phases at finite temperatures, and obtained its phase diagram with
respect to temperature and misfit strain. The phase diagram of SnTiO$_{3}$
turns out to be rather complicated, containing two triple points and
boundaries that separate ferroelectric phases. Such special features
provide unique opportunities to design novel high-performance ferroelectric
materials containing SnTiO$_{3}$, in which the rarely seen
$Cm$ phase can exist. In addition, while phonon calculation had shown
that AFD related modes are unstable \ucite{Parker2011}, no AFD-related
structural phases were found in our simulations, which is likely due
to the strong competition from unstable polar modes.

\end{document}